\documentclass[12pt]{article}

\usepackage[english]{babel}

\usepackage{amssymb,epsfig}
\usepackage{amsmath}

\usepackage{slashed}
\usepackage[active]{srcltx}
\usepackage{cite}
\usepackage{pdfsync}
\usepackage{epsfig,epsf}
\RequirePackage{graphicx}
\newcommand{\ms}{\mskip 1.5mu}

\newcommand{\Li}{\operatorname{Li}}

\newcommand \VEV [1] {\left\langle{#1}\right\rangle}

\newcommand{\beq}[1]{
\marginpar{\small\textsf{#1}}
\begin{equation}\label{#1}}

\newcommand{\eeq}{\end{equation}}

\newcommand{\bea}[1]{
\marginpar{\small\textsf{#1}}
\begin{eqnarray}\label{#1}}

\newcommand{\eea}{\end{eqnarray}}

\begin{document}


\begin{center}
\textbf{\LARGE  Two-loop evolution equations for light-ray operators}

\vspace*{1.6cm}

{\large V.M.~Braun$\ms{}^1$  and  A.N.~Manashov$\ms{}^{1,2}$ }

\vspace*{0.4cm}

\textsl{%
$^1$ Institut f\"ur Theoretische Physik, Universit\"at Regensburg,
D-93040 Regensburg, Germany \\
$^2$ Department of Theoretical Physics,  St.-Petersburg 
University, 199034, St.-Petersburg, Russia}

\vspace*{0.8cm}

\textbf{Abstract}\\[10pt]
\parbox[t]{0.9\textwidth}{
QCD in non-integer $d=4-2\epsilon$ space-time dimensions possesses a nontrivial critical point and
enjoys {\it exact} scale and conformal invariance. This symmetry imposes  nontrivial restrictions on the form of
the renormalization group equations for composite operators in physical (integer) dimensions and allows
to reconstruct full kernels from their eigenvalues (anomalous dimensions). We use this technique to derive
two-loop evolution equations for flavor-nonsinglet quark-antiquark light-ray operators that encode
the scale dependence of generalized hadron parton distributions and light-cone distribution amplitudes in
the most compact form.
}
\end{center}

\vspace*{1cm}



{\large\bf 1.}~~Studies of hard exclusive reactions contribute significantly to the research program at
all major existing and planned accelerator facilities. The relevant nonperturbative
input in such processes involves operator matrix elements between states with different
momenta, dubbed generalized parton distributions (GPDs), or vacuum-to-hadron matrix elements
related to light-front hadron wave functions at small transverse separations, the
distribution amplitudes (DAs). Scale dependence of these distributions is governed by the
renormalization group (RG) equations for the corresponding (nonlocal) operators and
has to be calculated to a sufficiently high order in perturbation theory in order to
make the QCD description of exclusive reactions fully quantitative.
At present, the evolution equations for GPDs (and DAs) are known to the two-loop
accuracy~\cite{Belitsky:1998vj,Belitsky:1999hf,Belitsky:1998gc},
one order less compared to the corresponding ``inclusive'' distributions that involve
forward matrix elements~\cite{Moch:2004pa,Vogt:2004mw} and closing this gap is desirable.
The direct calculation is very challenging, and also
finding a suitable representation for the results may become a problem as the
two-loop evolution equations for GPDs are already very cumbersome.

It has been known for some time~\cite{Mueller:1991gd} that conformal symmetry of the QCD Lagrangian allows one
to restore full evolution kernels at given order of perturbation theory from the spectrum of anomalous dimensions at the same order,
and the calculation of the special conformal anomaly at one order less.
This result was  used to calculate the complete two-loop mixing matrix for twist-two operators in
QCD~\cite{Mueller:1993hg,Mueller:1997ak,Belitsky:1997rh},
and derive the two-loop evolution kernels in momentum space for the
GPDs~\cite{Belitsky:1998vj,Belitsky:1999hf,Belitsky:1998gc}.
In Ref.~\cite{Braun:2013tva} we have suggested an alternative technique, the difference being that
instead of studying conformal symmetry \emph{breaking} in the physical theory~\cite{Mueller:1993hg,Mueller:1997ak,Belitsky:1997rh}
we make use of the \emph{exact} conformal symmetry of a modified theory -- QCD in $d=4-2\epsilon$ dimensions at critical coupling.
Exact conformal symmetry allows one to use algebraic group-theory methods to resolve the constraints on the
operator mixing and also suggests the optimal representation for the results in terms
of light-ray operators.
In this way a delicate procedure of the restoration of the evolution kernels
from the results for local operators is completely avoided.
We expect that these features will become increasingly advantageous in higher orders.

This modified approach was illustrated in~\cite{Braun:2013tva} on several examples to the two- and
three-loop accuracy for scalar theories. Application to gauge theories, in particular QCD,
involves several subtleties that are considered in this work. The main new result are the
two-loop evolution equations for flavor-nonsinglet quark-antiquark light-ray operators that encode
the scale dependence of generalized hadron parton distributions and light-cone distribution amplitudes in
the most compact form.

%
\vspace*{0.5cm}
%

{\large\bf 2.}~~
Before going over to technical details, let us first describe the general structure of the approach and the results
on a more qualitative level.
In order to make use of the (approximate) conformal symmetry of QCD it is natural to use a
coordinate-space representation in which the symmetry transformations have a simple form~\cite{Braun:2003rp}.
The relevant objects are light-ray operators that can be understood as generating functions for the
renormalized leading-twist local operators:
\begin{eqnarray}
 [\mathcal{O}](x;z_1,z_2) &\equiv& [\bar q(x+z_1n)\slashed{n} q(x+z_2n)]
~\equiv~ \sum_{m,k} \frac{z_1^m z_2^k}{m!k!} [(D_n^m\bar q)(x) \slashed{n} (D_n^k q)(x)].
\label{LRO}
\end{eqnarray}
Here $q(x)$ is a quark field, the Wilson line is implied between the quark fields on the light-cone,
$D_n = n_\mu D^\mu$ is a covariant derivative,  $n^\mu$ is an auxiliary light-like vector, $n^2=0$, that ensures
symmetrization and subtraction of traces of local operators.
The square brackets $[\ldots]$ stand for the renormalization using dimensional regularization and MS subtraction.
We will tacitly assume that the quark and antiquark have different flavor so that there is no mixing with gluon operators.
In most situations the overall coordinate $x^\mu$ is irrelevant and can be put to zero; we will often abbreviate
$\mathcal{O}(z_1,z_2) \equiv \mathcal{O}(0; z_1,z_2)$.

Light-ray operators satisfy a renormalization-group equation of the form~\cite{Balitsky:1987bk}
\begin{align}\label{RGO}
\Big(M{\partial_M}+\beta(g)\partial_g +\mathbb{H}\Big)[\mathcal{O}(z_1,z_2)]=0\,,
\end{align}
where $\mathbb{H}$ is an integral operator acting on the light-cone coordinates of the fields. It can be written as
\begin{align}
 \mathbb{H}[\mathcal{O}](z_1,z_2) = 
\int_0^1 \!d\alpha\int_0^1\! d\beta\, h(\alpha,\beta)\, [\mathcal{O}](z_{12}^\alpha,z_{21}^\beta)\,,
\label{hkernel}
\end{align}
where 
\begin{align}
z_{12}^\alpha\equiv z_1\bar\alpha+z_2\alpha\,, && \bar\alpha=1-\alpha\,,
\end{align}
and $h(\alpha,\beta)$ is a certain weight function (kernel).

One can show, see e.g.~\cite{Braun:2013tva}, that the powers $[\mathcal{O}](z_1,z_2)  \mapsto (z_{1}-z_{2})^N$ are
eigenfunctions of the evolution operator
$\mathbb{H}$, and the corresponding eigenvalues
\begin{align}\label{hmoments}
\gamma_N=\int d\alpha d\beta\, h(\alpha,\beta)(1-\alpha-\beta)^{N-1}\,
\end{align}
are nothing else as the anomalous dimensions of local operators of spin $N$ (with $N-1$ derivatives).

In general the function $h(\alpha,\beta)$ is a function of two variables and therefore the knowledge
of the anomalous dimensions $\gamma_N$ is not sufficient to fix it.
However, if the theory is conformally invariant then $\mathbb{H}$ must commute with the generators of the
$SL(2)$ transformations $[\mathbb{H},S_\alpha^{(0)}]=0$, where
\begin{align}
S^{(0)}_+=z_1^2\partial_{z_1}+z_2^2\partial_{z_2}+2(z_1+z_2),&&
S^{(0)}_0=z_1\partial_{z_1}+z_2\partial_{z_2}+2, &&
S^{(0)}_-=-\partial_{z_1}-\partial_{z_2}\,.
\label{canon}
\end{align}
In this case it can be shown that the function $h(\alpha,\beta)$ 
(up to trivial terms $\sim\delta(\alpha)\delta(\beta)$ that correspond to the unit operator) 
takes the form~\cite{Braun:1999te}
\begin{align}\label{hinv}
h (\alpha,\beta) = \bar h (\tau)\,, && \tau = \frac{\alpha\beta}{\bar\alpha\bar\beta}\,
\end{align}
and is effectively a function of one variable $\tau$ called the conformal ratio.
This function can easily be reconstructed from its moments (\ref{hmoments}), alias from the anomalous dimensions.

Conformal symmetry of QCD is broken by quantum corrections which implies that the symmetry
of the evolution equations is lost at the two-loop level.
In other words, writing the evolution  kernel as an expansion in the coupling constant
\begin{align}
  \mathbb{H} = a_s\,\mathbb{H}^{(1)} + a^2_s\,\mathbb{H}^{(2)}+\ldots
\qquad\mapsto\qquad
  h(\alpha,\beta) = a_s\, h^{(1)}(\alpha,\beta) + a^2_s\, h^{(2)}(\alpha,\beta)+\ldots,
\label{Hexpansion}
\end{align}
where $a_s= \alpha_s/(4\pi)$,
we expect that $h^{(1)}(\alpha,\beta)$ only depends on the conformal ratio whereas higher-order contributions remain to
be nontrivial functions of two variables $\alpha$ and $\beta$.

This prediction is confirmed by the explicit calculation~\cite{Balitsky:1987bk}:
\begin{align}
\mathbb{H}^{(1)}f(z_1,z_2)&=4C_F\biggl\{
\int_0^1d\alpha\frac{\bar\alpha}{\alpha}\Big[2f(z_1,z_2)-f(z_{12}^\alpha,z_2)-f(z_1,z_{21}^\beta)\Big]
\notag\\
&\quad
-\int_0^1d\alpha\int_0^{\bar\alpha}d\beta \, f(z_{12}^\alpha,z_{21}^\beta)
+\frac12 f(z_1,z_2)
\biggr\}\,.
\end{align}
The corresponding one-loop kernel $h^{(1)}(\alpha,\beta)$ can be written in the following,
remarkably simple form~\cite{Braun:1999te}
\begin{align}
      h^{(1)}(\alpha,\beta) = -4 C_F\left[\delta_+(\tau) + \theta(1-\tau)-\frac12\delta(\alpha)\delta(\beta)\right],
\label{QCD-LO}
\end{align}
where the regularized $\delta$-function, $\delta_+(\tau)$, is defined as
\begin{align}
\int d\alpha d\beta\, \delta_+(\tau)f(z_{12}^\alpha,z_{21}^\beta)&\equiv\int_0^1 d\alpha\int_0^{1} d\beta\, \delta(\tau)
\Big[f(z_{12}^\alpha,z_{21}^\beta)-f(z_1,z_2)\Big]
\notag\\
&=-
\int_0^1 d\alpha \frac{\bar \alpha}{\alpha}\Big[2f(z_1,z_2)-f(z_{12}^\alpha,z_2)-f(z_1,z_{21}^{\alpha})\Big].
\label{delta-function}
\end{align}
Taking appropriate matrix elements and making a Fourier transformation to the momentum fraction
space one can check that the expression in Eq.~(\ref{QCD-LO}) reproduces all classical 
leading-order (LO) QCD evolution equations: DGLAP equation for parton distributions, 
ERBL equation for the meson light-cone DAs, and the general evolution equation for GPDs.

The two-loop kernel $h^{(2)}(\alpha,\beta)$ contains contributions of two color structures and a term proportional to the
QCD beta function,
\begin{align}\label{CACFH}
  h^{(2)}(\alpha,\beta) =8 C_F^2  h^{(2)}_1(\alpha,\beta) +4 C_F C_A h^{(2)}_2(\alpha,\beta) +4 b_0 C_F h^{(2)}_3(\alpha,\beta)\,.
\end{align}

Let us explain how it can be calculated. The idea of Ref.~\cite{Braun:2013tva} is to consider a modified theory, QCD
in non-integer $d=4-2\epsilon$ dimensions. In this theory the $\beta$-function has the form
\begin{align}
\beta(a)=M\partial_M a=2a\Big(-\epsilon - b_0 a+ \mathcal{O}(a^2)\Big)\,, && b_0 = \frac{11}{3} N_c - \frac23 n_f\,,
\end{align}
and for a large number of flavors $n_f$ there exists a critical coupling $a^\ast_s= \alpha_s^*/(4\pi) \sim \epsilon$ such that
$\beta(a_s^{*})=0$.  The theory thus enjoys exact scale invariance~\cite{Banks:1981nn,Hasenfratz:1992jv}
and one can argue (see below) that full conformal invariance is also present.%
\footnote{Formally the gauge-fixed QCD Lagrangian contains two charges, the coupling and the gauge parameter. The corresponding
$\beta$-function, $\beta_\xi = M\partial_M \xi$, vanishes in the Landau gauge, $\xi=0$,
so that {\it all} Green functions are scale-invariant at the critical point in this gauge;
$\beta_\xi$ also drops out of the RG equations for the correlation functions
of gauge-invariant operators.}
As a consequence, the renormalization group equations are exactly conformally invariant: the
evolution kernels commute with the generators of the conformal group.
The generators are, however, modified by quantum corrections as compared to their canonical
expressions (\ref{canon}):
\begin{align}
 S_\alpha=S_\alpha^{(0)}+a_s^\ast\,S^{(1)}_\alpha + (a_s^\ast)^2 \, S^{(2)}_\alpha +\ldots
\label{expandS}
\end{align}
 One can show that
\begin{align}
   S_- =& S_-^{(0)}\,,
\notag\\
   S_0\, =& S_0^{(0)} -\epsilon+\frac12 \mathbb{H}(a_s^*)\,, \qquad \mathbb{H}(a_s^*) = a^\ast_s\,\mathbb{H}^{(1)}+\ldots
\notag\\
   S_+ =& S_+^{(0)} + (z_1+z_2)\Big(-\epsilon+ \frac12 a_s^\ast \mathbb{H}^{(1)}\Big)
+  a_s^\ast (z_1-z_2)\Delta_+ + \mathcal{O}(\epsilon^2)\,,
\label{exactS}
\end{align}
where
\begin{align}
\Delta_+ [\mathcal{O}](z_1,z_2)
= -2C_F\int_0^1d\alpha\Big(\frac{\bar\alpha}\alpha+\ln\alpha\Big) \Big[[\mathcal{O}](z_{12}^\alpha,z_2)-[\mathcal{O}](z_1,z_{21}^\alpha)\Big]
\label{result1}
\end{align}
i.e. the generator $S_-$ is not modified, the deformation of $S_0$ can be calculated exactly in
terms of the evolution operator (to all orders in perturbation theory)~\cite{Braun:2013tva},
whereas the deformation of $S_+$ is nontrivial and has to be calculated explicitly order by order, 
to the required accuracy.
The one-loop expression shown in (\ref{exactS}), (\ref{result1}) is derived below, it is a new result.
From the pure technical point of view, this calculation replaces evaluation of the conformal
anomaly in the theory with broken symmetry in integer dimensions via the conformal Ward identities (CWI)
in the approach due to D.~M{\"u}ller~\cite{Mueller:1991gd}.

Conformal symmetry of the modified QCD at the critical coupling implies that the generators (\ref{exactS}) satisfy the
usual $SL(2)$ commutation relations
\begin{align}\label{sl2-comm}
{}[S_0,S_{\pm}]=\pm S_{\pm}\,, &&
{}[S_{+},S_-]= 2S_0\,.
\end{align}
Expanding them in powers of the coupling $a_s^\ast$ one obtains a nested set of commutator relations~\cite{Braun:2013tva}
\begin{align}
{}[S_+^{(0)},\mathbb{H}^{(1)}]=&~0\,,
\notag\\
{}[S_+^{(0)},\mathbb{H}^{(2)}]=&~[\mathbb{H}^{(1)},\Delta S_+^{(1)}]\,,
\notag\\
{}[S_+^{(0)},\mathbb{H}^{(3)}]=&~[\mathbb{H}^{(1)},\Delta S_+^{(2)}]+[\mathbb{H}^{(2)},\Delta S_+^{(1)}]\,,
\label{nest}
\end{align}
etc. Note that the commutator of the canonical generator $S_+^{(0)}$ with the evolution kernel at order $k$
on the l.h.s. of each equation 
is given in terms of the evolution kernels $\mathbb{H}^{(k)}$ and the corrections to the generators
$\Delta S_+^{(m)}$ at one order less, $m \le k-1$.
The commutation relations  Eq.~(\ref{nest}) can be viewed as, essentially, inhomogeneous first-order
differential equations on the evolution kernels. Their solution determines $\mathbb{H}^{(k)}$ up to an
$SL(2)$-invariant term (solution of a homogeneous equation  $[\mathbb{H}_{inv}^{(k)},S_\alpha^{(0)}]=0$),
which can, again, be restored from the spectrum of the anomalous dimensions. This procedure is described
in detail for scalar theories in Ref.~\cite{Braun:2013tva}.

Last but not least, in $\text{MS}$-like schemes the evolution kernels (anomalous dimensions)
do not depend on the space-time dimension by construction.
Indeed, the renormalization $\mathbb{Z}$ factors relating the renormalized and bare light-ray operators
$[\mathcal{O}](z_1,z_2) = \mathbb{Z}\,\mathcal{O}(z_1,z_2)$ are given by the expansion
\begin{align}
\mathbb{Z}=1+\sum_{j=1}^\infty \epsilon^{-j}\sum_{k=j}^\infty  a_s^k \, \mathbb{Z}_{jk}\,,
\end{align}
where $\mathbb{Z}_{jk}$ have the integral representation similar to (\ref{hkernel}) in terms of functions
of two variables, $Z_{jk}(\alpha,\beta)$ that do not depend on $\epsilon$.
Thus, eliminating the $\epsilon$-dependence of the expressions derived in the $d$-dimensional
(conformal) theory for the critical coupling by the expansion $\epsilon = -b_0 a_s^\ast + \mathcal{O}(a^{*2}_s)$
allows one to restore the evolution kernels for the theory in integer dimensions for arbitrary coupling  $a_s^\ast\to a_s$;
this rewriting is simple and exact to all orders.

%
\vspace*{0.5cm}
%

{\large\bf 3.}~~
The statement of conformal invariance of QCD in $d$ dimensions at the critical coupling is not trivial.
It is  believed that ``physically reasonable'' scale-invariant theories
are also conformally invariant, see Ref.~\cite{Nakayama:2013is} for a discussion, however,
to the best of our knowledge there is no proof of this statement for $d>2$ dimensions
(but there are no counterexamples as well). In non-gauge theories conformal invariance for the Green functions
of basic fields can be checked in perturbative expansions~\cite{Derkachov:1993uw}.
In gauge theories  conformal invariance does not hold for the correlators of  basic fields and
can be expected only for the Green functions of gauge-invariant operators.
For local composite operators a proof of conformal invariance is based on the analysis
of pair counterterms for the product of the trace of energy-momentum tensor and local operators~\cite{AN}.
This analysis is beyond the scope of this Letter; it becomes rather complicated in 
gauge theories due to mixing of gauge invariant operators with BRST
variations and equation of motion (EOM) operators~\cite{Collins}.

A  short comment may, nevertheless, be relevant.
Let $\mathcal{O}_N$ be a gauge-invariant multiplicatively renormalizable operator
\begin{align}\label{mRGE}
\Big(M\partial_M+\beta(a)\partial_a+\gamma_N(a)\Big)[\mathcal{O}_N]=0\,,
\end{align}
where 
$\gamma_N(a)$ is the anomalous dimension.
%
 As a consequence, it possesses a certain (critical)
dimension for the fine-tuned value of the coupling (critical point) $a=a_*$, $\beta(a_*)=0$:
\begin{align}\label{DO}
i [\mathbf{D},[\mathcal{O}_{N}](x)] =\Big(x\partial_x+\Delta_{N}^*\Big)\,[\mathcal{O}_{N}](x)\,,
\end{align}
where $\mathbf{D}$ is the operator of dilatations,
$\Delta_{N}$ is the canonical dimension of the operator $\mathcal{O}_{N}$,
and $\Delta_{N}^*=\Delta_{N}+\gamma_N^*$ is the scaling dimension, $\gamma_N^*=\gamma_N(a_*)$.

The statement that $[\mathcal{O}_{N}](x)$ becomes a conformal operator at the critical point,
as widely expected, means that action of the generator of special conformal transformations
on this operator takes the form
\begin{align}\label{SKO}
i[\mathbf{K}^\mu,[\mathcal{O}_{N}](x)]=\left[2x^\mu(x\partial)-x^2\partial^\mu+2\Delta_N^* x^\mu
+2x^\nu\left(n^\mu\frac{\partial}{\partial n^\nu}-n_\nu\frac{\partial}{\partial n_\mu}\right)\right]\,[\mathcal{O}_{N}](x)\,.
\end{align}
Equivalently, a correlation function of such operators at the critical point must satisfy the Ward identity
\begin{align}\label{confI}
\left( K^{\mu}_{x_1}+\ldots+ K_{x_n}^{\mu}\right)\VEV{[\mathcal{O}_1](x_1)\ldots [\mathcal{O}_n](x_n)}=0\,,
\end{align}
where it is assumed that all space-time points $x_i$ are different.
Calculating the l.h.s. in perturbation theory (see Ref.~\cite{AN})
and making use of the dilatation Ward identity produces the expression of the form
\begin{align}
\left( K^{\mu}_{x_1}+\ldots+ K_{x_n}^{\mu}\right)\VEV{[\mathcal{O}_1](x_1)\ldots [\mathcal{O}_n](x_n)}
=\sum_{i=1}^N \VEV{[\mathcal{O}_1](x_1)\ldots \widetilde{ \mathcal{O}}_i^\mu(x_i)\ldots[\mathcal{O}_n](x_n)},
\label{confII}
\end{align}
where $\widetilde{ \mathcal{O}}_i(x_i)$ are local operator insertions that involve several contributions:
EOM operators, operators representing a BRST variation of another operator,
and gauge-invariant operators. The first two can be neglected as the do not contribute to the correlation function
(assuming $x_j\neq x_k$, for all $j\neq k$). The last ones can further be expanded in terms of
gauge invariant operators $[\mathcal{O}^\mu_{iq}](x_i)$ with a certain critical dimension,
$$\widetilde{ \mathcal{O}}_i^\mu(x_i)=\sum_q c_q(a) [\mathcal{O}^\mu_{iq}](x_i).$$
Dilatation invariance implies that the both sides of Eq.~(\ref{confII}) must have the same
scaling dimension at the critical point.
Note that application of $K^{\mu}$ lowers the scaling dimension of an operator by one.
As a consequence, for each contribution on the r.h.s., either $c_q(a_*)$ vanishes, or the
scaling dimensions of $[\mathcal{O}^\mu_{iq}]$ and $[\mathcal{O}_i]$ must differ by one,
$\dim\mathcal{O}^\mu_{iq}=\dim \mathcal{O}_i-1$, to all orders of perturbation theory.
If such operators do not exist, then all coefficient $c_q(a)$ have to vanish at the critical point that
implies conformal invariance Eq.~(\ref{confI}).
For the leading-twist operators that  are subject of this Letter, absence of operators with the same anomalous
dimension and the canonical dimension less by one is easy to verify.
For the general case, it would be quite unexpected that
two different operators have the same anomalous dimension if they are not related by some exact symmetry,
although existence of such pairs cannot be excluded.

%
\vspace*{0.5cm}
%

{\large\bf 4.}~~
The correction $\Delta S_+$, $S_+ = S_+^{(0)} +\Delta S_+$, 
to the generator of special conformal transformations at the critical point 
in QCD in $d$ dimensions (in the light-ray operator representation) 
can be derived from the analysis of CWIs for suitable correlation functions.
The standard procedure is to consider the Green functions of twist-two operators
with quark fields, see Refs.~\cite{Mueller:1993hg,Belitsky:1998vj} (for $d=4$),
which are not gauge invariant that complicates the analysis.
This difficulty can be avoided by considering, instead,
the correlator of two light-ray operators that depend on
different auxiliary vectors $n$ and $\bar n$:
\begin{align}\label{G-int}
G(x, z, w) & = \VEV{[\mathcal{O}^{(n)}](0,z)\,[\mathcal{O}^{(\bar n)}](x,w)}
 ~=~\mathcal{N}^{-1}\int D\Phi\, e^{-S_R(\Phi)}\,[\mathcal{O}^{(n)}](0,z)\,[\mathcal{O}^{(\bar n)}](x,w)\,,
\end{align}
where $\Phi\equiv \{A, q,\bar q, c,\bar c\}$ is the set of fundamental fields,
$z \equiv \{z_1,z_2\}$, $w\equiv \{w_1,w_2\}$ and we assume that the auxiliary light-like vectors are normalized as
$(n\cdot\bar n)=1$.

The QCD Lagrangian in $d = 4 - 2\epsilon$ dimensional Euclidean space-time in covariant gauge has the
form%
\footnote{Our notation follows closely Ref.~\cite{Ciuchini:1999cv}}
\begin{align}\label{QCD-L}
\mathcal{L}=\bar q(\slashed{\partial}-ig\slashed{A})q+\frac14 F_{\mu\nu}^a F^{a,\mu\nu}+
\partial_\mu \bar c^a(D^\mu c)^a+\frac1{2\xi}(\partial A^a)^2.
\end{align}
 The renormalized action $S_R$ is obtained from~(\ref{QCD-L}) by the replacement
\begin{align}
\Phi\to \Phi_0=Z_\Phi \Phi, && g\to g_0=M^{\epsilon} Z_g g\,, &&\xi\to
\xi_0 =Z_\xi \xi\,,
\end{align}
i.e.~$S_R(\Phi,g,\xi)= S(\Phi_0, g_0,\xi_0)$.
Note that we do not send $\epsilon\to 0$ in the action and the renormalized correlation functions so that
they explicitly depend on $\epsilon$.

The form of the CWI is simpler for the special choice $(n\cdot x)=(\bar n\cdot x)=0$ that we accept for this calculation.
For the local conformal operators defined with respect to the $\bar n$ vector, $\mathcal{O}_N^{(\bar n)}(x)$,
cf.~\cite{Braun:2003rp}, it follows from from Eq.~(\ref{SKO}) that
\begin{align}
i[(\bar n \mathbf{K}), \mathcal{O}_N^{(\bar n)}(x)]=-x^2(\bar n\partial_x)\,\mathcal{O}_N^{(\bar n)}(x)\,.
\end{align}
Going over to the light-ray operators one obtains, therefore
\begin{align}
  i[(\bar n \mathbf{K}), \mathcal{O}^{(\bar n)}(x,w)]=-x^2(\bar n\partial_x)\,\mathcal{O}^{(\bar n)}(x,w)\,.
\end{align} 
Thus conformal invariance of the correlation function (\ref{G-int}) at the critical point implies the
constraint
\begin{eqnarray}
\lefteqn{\hspace*{-1cm}
\frac{i}{2} \VEV{
[(\bar n \mathbf{K}),\mathcal{O}^{(n)}](0,z)\,\mathcal{O}^{(\bar n)}(x,w)
+\mathcal{O}^{(n)}(0,z)\,(\bar n \mathbf{K}),\mathcal{O}^{(\bar n)}(x,w)]
                } =}
\nonumber\\&&{} \hspace*{5cm}{}=\,
\Big(S_+^{(z)}-\frac12x^2(\bar n\partial_x)\Big)G(x,{z},{w})=0\,,
\end{eqnarray}
where the superscript $S_+^{(z)}$ reminds that it is a differential operator acting on $z_1,z_2$ coordinates.

Explicit expression for $S_+^{(z)}$ can be derived from the CWI
\begin{align}\label{OOS}
0  &=-\VEV{\delta_+ S_R\, [\mathcal{O}^{(n)}](z)\,[\mathcal{O}^{(\bar n)}](x,w)}
+
\VEV{\delta_+[\mathcal{O}^{(n)}](z)\,[\mathcal{O}^{(\bar n)}](x,w)}
+\VEV{[\mathcal{O}^{(n)}](z)\,\delta_+[\mathcal{O}^{(\bar n)}](x,w)}\,,
%
\end{align}
where $[\mathcal{O}](z_1,z_2)\equiv[\mathcal{O}](x=0;z_1,z_2)$,
that follows from invariance of the correlation function~(\ref{G-int}) under a change of variables
$\Phi\mapsto \Phi+\delta_+ \Phi$ in the path integral,
\begin{align}\label{Conf}
\delta_+ q(x)&=\bar n^\mu\Big(
\big(2x_\mu(x\partial)-x^2\partial_\mu +2\Delta_q \,x_\mu\big) q(x)+\frac12[\gamma_\mu,\slashed{x}]q(x)\Big)\,,
\notag\\
\delta_+ A_\rho(x)&=\bar n^\mu\Big(\big(2x_\mu(x\partial)-x^2\partial_\mu +2\Delta_A \,x_\mu\big) A_\rho(x)
+2g_{\mu\rho} (x A)-2x_\rho A_\mu(x)\Big)\,,
\notag\\
\delta_+ c(x)&=\bar n^\mu\Big(2x_\mu(x\partial)-x^2\partial_\mu +2\Delta_c\, x_\mu\Big) c(x)\,,
\notag\\
\delta_+ \bar c(x)&=\bar n^\mu\Big(2x_\mu(x\partial)-x^2\partial_\mu +2\Delta_{\bar c}\, x_\mu\Big) \bar c(x)\,.
\end{align}
The  choice of the parameters $\Delta_\Phi$ is a matter of convenience. They can be taken, e.g., equal to the canonical
dimensions of the fields in $d$ space-time dimensions, as in~\cite{Braun:2013tva}. For QCD a different choice proves to
be more convenient: $\Delta_q=3/2-\epsilon$,  $\Delta_A=1$, $\Delta_{\bar c}=2$ and $\Delta_c=0$.
In this case the quark part of the Lagrangian is invariant and variation of the action takes the form
\begin{align}\label{KS}
\delta_+ S_R =4\epsilon\int d^dx
(x\bar n) (\mathcal{L}_A+\mathcal{L}_\xi+\mathcal{L}_{ghost})
+2(d-2)\bar n^\mu\int d^dx \Big(Z_c^2\,\bar c \,D_\mu c-\frac1\xi A_\mu (\partial A)\Big).
\end{align}
The reason for choosing different scaling dimensions for the ghost and anti-ghost fields 
is that in this case   
the last term $\sim (d-2)$ that does not vanish in four dimensions is a BRST
variation~\cite{Belitsky:1999hf}
\begin{align}\label{BRST}
\Big(Z_c^2\bar c^a D_\mu c^a-\frac1\xi A_\mu^a (\partial A^a)\Big)=\delta_{BRST} (\bar c^{a} A_\mu^a)\,.
\end{align}
Hence, it does not  contribute to the variation of~(\ref{G-int}).

In this work we are interested in the one-loop correction to the generator $S_+$. To this accuracy, obviously,
the ghost Lagrangian $\mathcal{L}_{ghost}$ and gluon self-interaction do not contribute i.e. we have to keep terms
quadratic in gluon fields only. One obtains after some algebra
\begin{align}\label{SK2}
\delta_+ S_R=-2\epsilon \int d^dx \left[(x\bar n) \, A_\alpha^a\, K^{\alpha\beta} A^a_\beta+
\Big(1+\frac1\xi\Big)(\bar n A^a) (\partial A^a)
\right]+\ldots\,,
\end{align}
where the ellipses stands for the terms that are irrelevant at one loop order and
\begin{align}
 K^{\alpha\beta} =g^{\alpha\beta}\partial^2-\partial^\alpha\partial^\beta\left(1-\frac1\xi\right)\,
\end{align}
is nothing but the inverse gluon propagator (with a minus sign).
Moreover, as follows from Eq.~(\ref{BRST}), the last term $\sim (\bar n A) (\partial A)$ can be written as
a combination of the BRST variation and the ghost term, so that it does not contribute to the one-loop accuracy as well.
Thus, to our accuracy, the insertion of $\delta_+\, S_R$ generates an effective vertex insertion
$-2\epsilon\,(\bar n x) \, A_\alpha^a(x)\, K^{\alpha\beta} A^a_\beta(x)$ in a gluon line in one loop diagrams:
\begin{align}\label{eff-prop}
\begin{minipage}{3.2cm}{\includegraphics[width=3.2cm]{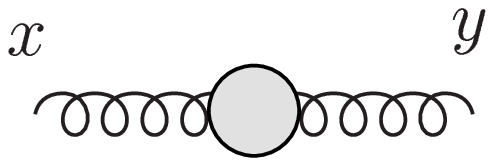}}\\ \end{minipage}
\begin{minipage}{2.6cm}{$=2\epsilon \, \bar n\cdot(x+y)$}\end{minipage}
\begin{minipage}{3cm}{\includegraphics[width=3cm]{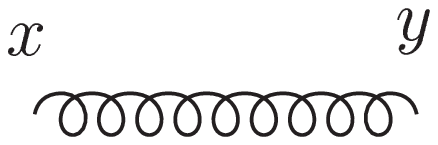}\\}\end{minipage}
\end{align}
%
%

%
\begin{figure}[t]
\begin{center}
\includegraphics[width=.850\textwidth]{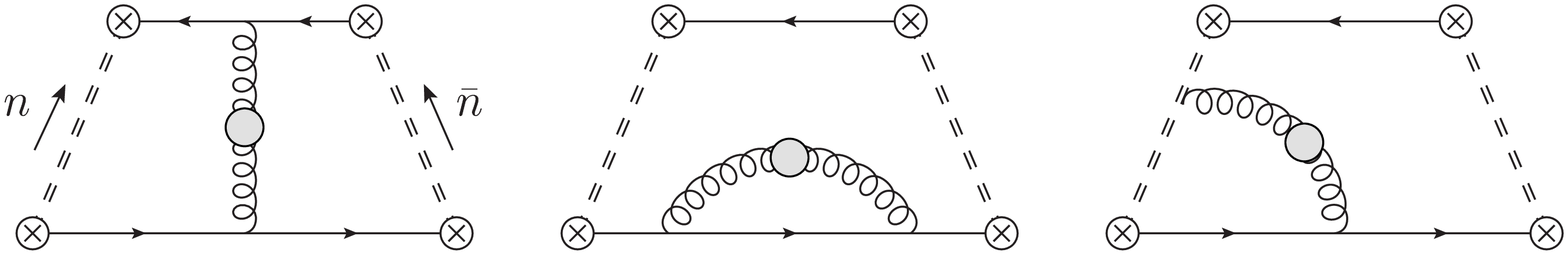}
\end{center}
\caption{One-loop diagrams contributing to the correlation function of two light-ray operators with an insertion
of $\delta_+S_R$ (grey blobs).
Wilson lines between the quark fields are shown by the dashed double lines.
}
\label{fig1}
\end{figure}
%

Feynman diagrams contributing to the correlation function $\VEV{\delta_+S_R [\mathcal{O}^{(n)}](z)[\mathcal{O}^{(\bar n)}](x,w)}$
on the r.h.s. of the CWI~(\ref{OOS}) are shown in Fig.~\ref{fig1}.
Since the $\delta_+S_R$ insertion brings the factor $\epsilon g^2$, we only need the divergent parts.
The calculation is very easy. For definiteness let us choose the Feynman gauge, $\xi=1$,
and consider the first diagram in Fig.~\ref{fig1} as an example.
This diagram, potentially, has two divergent subgraphs --- left and right --- but the right one
does not contribute due to our choice $(\bar n, x)=0$, taking into account that the operator
$\mathcal{O}^{(\bar n)}(x,w)$ only involves quark fields along the line $\bar n^\mu$.
Thanks to (\ref{eff-prop}) the divergent part of the left subgraph coincides up to a prefactor $2\epsilon (\bar n,y+y')$,
with the corresponding counterterm for the operator  $\mathcal{O}^{(n)}(z_1,z_2)$, which is known from Ref.~\cite{Balitsky:1987bk}.
One obtains for this contribution
\begin{align}
(n\bar n)\frac{\alpha_s}{\pi} C_F \mathcal{H}^{(+)} (z_1+z_2)\,{G}^{(0)}(x,{z},{w})\,,
\end{align}
where ${G}^{(0)}(x,{z},{w})$ is the tree-level correlation function (\ref{G-int}) and
the integral operator $ \mathcal{H}^{(+)}$ is defined as follows:
\begin{align}
 \mathcal{H}^{(+)} f(z_1,z_2)=\int_0^1 d\alpha\int_0^{\bar\alpha}d\beta\, f(z_{12}^\alpha, z_{21}^\beta)\,.
\end{align}
The other diagrams in Fig.~\ref{fig1}, similarly, are written in terms of contributions of the
corresponding counterterms to $\mathcal{O}^{(n)}(z_1,z_2)$ decorated by multiplicative factors.
For the sum of all terms one obtains
\begin{eqnarray}
\lefteqn{\VEV{\delta_+ S_R\, [\mathcal{O}^{(n)}](0,z)\,[\mathcal{O}^{(\bar n)}](x,w)}=}
\\ &=&
(n\bar n)\frac{\alpha_s}{\pi} C_F
\Big[-\{\widehat{\mathcal{H}}_1, z_1\}-\{\widehat{\mathcal{H}}_2, z_2\}
+
\left(\mathcal{H}^{(+)}-\frac12\right)(z_1\!+\!z_2)+z_{12} \,\widehat{\mathcal{H}}'\Big]{G}^{(0)}(x,{z},{w})
+\mathcal{O}(\epsilon)
\,,
\nonumber
\end{eqnarray}
where $\{*,*\}$ stands for an  anticommutator and
\begin{align}
\widehat{\mathcal{H}}_1 f(z_1,z_2)&=\int_0^1d\alpha\frac{\bar\alpha}{\alpha}\,\big[f(z_1,z_2)-f(z_{12}^\alpha,z_2)\big],
\notag\\
\widehat{\mathcal{H}}_2 f(z_1,z_2)&=\int_0^1d\alpha\frac{\bar\alpha}{\alpha}\,\big[f(z_1,z_2)-f(z_1,z_{21}^\alpha)\big],
\notag\\
\widehat{\mathcal{H}}'f(z_1,z_2)&=\int_0^1d\alpha\ln\alpha\,\big[f(z_{12}^\alpha,z_2)-f(z_1,z_{21}^\alpha)\big].
\end{align}
The one-loop evolution operator $\mathbb{H}^{(1)}$ in Eq.~(\ref{Hexpansion}) is written in terms of these
kernels as
\begin{align}
\mathbb{H}^{(1)}=4C_F\Big[\widehat{\mathcal{H}}_1+\widehat{\mathcal{H}}_2-\mathcal{H}^{(+)}+\frac12\Big].
\end{align}
This representation is equivalent to the one in Eq.~(\ref{QCD-LO}) and, as it is easy to show,
$\mathbb{H}^{(1)}$ commutes with $S_+^{(0)}$.

Next, we have to consider the second contribution on the r.h.s. of the CWI~(\ref{OOS}), which involves
the conformal variation of $[\mathcal{O}^{(\bar n)}](x,w)$,
\begin{align}
\delta_+[\mathcal{O}^{(\bar n)}](x,w)=\mathbb{Z}\delta_+ \mathcal{O}^{(\bar n)}(x,w)
=\mathbb{Z}\big(-x^2(\bar n\partial_x)\big)\mathcal{O}^{(\bar n)}(x,w)
=-x^2(\bar n\partial_x)[\mathcal{O}^{(\bar n)}](x,w)\,,
\end{align}
and, finally, the third contribution
\begin{align}\label{dSO}
\delta_+[\mathcal{O}^{(n)}](z)&=
  2(n\bar n)\mathbb{Z}\delta_+ [\mathcal{O}^{(n)}](z)
= 2(n\bar n)\mathbb{Z}\Big(S_+^{(0)}-\epsilon(z_1+z_2)\Big)\mathbb{Z}^{-1}[\mathcal{O}^{(n)}](z)
\notag\\
&=2(n\bar n)
\Big(S_+^{(0)}-\epsilon(z_1+z_2)-\frac{a_s}{2}[\mathbb{H}^{(1)},z_1+z_2)]\Big)[\mathcal{O}^{(n)}](z)\,.
\end{align}
This last contribution is discussed in detail in Ref.~\cite{Braun:2013tva} where the chain of equations
in (\ref{dSO}) is explained.
Collecting all terms we obtain the result quoted in Eq.~(\ref{exactS}).
Note that this expression is different from the corresponding result in scalar field theories,
$$S_+=S_+^{(0)}+(z_1+z_2)\Big(-\epsilon+\frac12a_*\mathbb{H}^{(1)}\Big),$$
see Ref.~\cite{Braun:2013tva}.

%
\vspace*{0.5cm}
%

{\large\bf 5.}~~
We proceed to calculate the NLO evolution kernels (\ref{CACFH}) making use of
the commutator relation $[S_+^{(0)},\mathbb{H}^{(2)}] = [\mathbb{H}^{(1)},\Delta S_+^{(1)}]$, Eq.~(\ref{nest}).
Note that $\Delta S_+^{(1)}$ (\ref{exactS}) contains terms in $b_0$ and $C_F$%
~\footnote{To the one-loop accuracy one can replace $\epsilon = (4-d/2)$ by $-b_0 a_s^*$.}.
Hence the commutator $[\Delta S_+^{(1)},\mathbb{H}^{(1)}]$ contains two color structures, $b_0 C_F$ and $C_F^2$,
respectively. It follows that the kernel $C_FC_A h^{(2)}_2(\alpha,\beta)$~(\ref{CACFH}) satisfies the homogeneous
equation $[S_+^{(0)},\mathbb{H}_2^{(2)}] = 0 $, alias it is $SL(2)$-invariant and can be written as a function of the
conformal ratio, $h^{(2)}_2(\alpha,\beta)= h^{(2)}_2(\tau)$.

Calculating the commutator we obtain after some algebra
\begin{align}
{}[\Delta S_+^{(1)},\mathbb{H}^{(1)}]=8\,C_F^2\,\mathbb{A}_1 + 4 \,C_F\, b_0\, \mathbb{A}_3\,,
\end{align}
where $\mathbb{A}_1$ and  $\mathbb{A}_3$ are integral operators of the form
\begin{align*}
\mathbb{A}_i f(z_1,z_2)=z_{12}\int_0^1 d\alpha\, A_i(\alpha) \big[f(z_{12}^\alpha,z_2)-f(z_1,z_{21}^\alpha)\big]
+z_{12} \int_0^1d\alpha\int_0^{\bar\alpha}d\beta\, B_i(\alpha,\beta)\, f(z_{12}^\alpha,z_{21}^\beta)\,,
\end{align*}
with
\begin{align}
A_1(\alpha)&=\ln\alpha\ln\bar\alpha-\frac32\bar\alpha-(2-\alpha)\ln\alpha+\frac{2+\alpha}\alpha\bar\alpha\ln\bar\alpha\,,
\notag\\
B_1(\alpha,\beta)&=(\beta-\alpha)\Big[\frac32-\ln(1-\alpha-\beta)\Big]
-\alpha\ln\alpha+\beta\ln\beta+\frac1\alpha\ln\bar\alpha-\frac1\beta\ln\bar\beta\,,
\notag\\
A_3(\alpha)&=\bar\alpha\,, \qquad B_3(\alpha,\beta)=\alpha-\beta\,.
\end{align}
The simplest way to calculate the evolution kernels $h^{(2)}_{1}$, $h^{(2)}_{3}$ is to try the 
following ansatz:
%
\begin{align}\label{ansatz}
h^{(2)}_{k}(\alpha,\beta)=-\delta_+(\tau)\Big[\phi_k(\alpha)+\phi_k(\beta)\Big]+\varphi_k(\alpha,\beta)+c_k\delta(\alpha)\delta(\beta)\,.
\end{align}
Calculating the commutator $[S_+^{(0)},\mathbb{H}^{(2)}]$ one obtains first-order differential equations on the functions
$\phi_k,\varphi_k$ (the terms in $c_k$ drop out from the commutator)
\begin{align}
\bar\alpha^2\partial_\alpha \phi_k(\alpha)=-A_k(\alpha)\,,
&&
\left(\alpha\bar\alpha\partial_\alpha-\beta\bar\beta\partial_\beta\right)\varphi_k(\alpha,\beta)=B_k(\alpha,\beta)\,,
\end{align}
where $\partial_\alpha = d/d\alpha$, etc. 
The solutions can be chosen as
\begin{align}\label{phi1}
\phi_1(\alpha)&=-\ln\bar\alpha\left[\frac32-\ln\bar\alpha
+\frac{1+\bar\alpha}{\bar\alpha}\ln\alpha\right], \qquad \phi_3(\alpha) =\ln\bar\alpha\,,
\notag\\
\varphi_1(\alpha,\beta)&
=-\theta(1-\tau)\Big[\frac12\ln^2(1-\alpha-\beta)+\frac12\ln^2\bar\alpha+\frac12\ln^2\bar\beta-\ln\alpha\ln\bar\alpha-\ln\beta\ln\bar\beta
\notag\\
&\quad-\frac12\ln\alpha-\frac12\ln\beta
+\frac{\bar\alpha}\alpha\ln\bar\alpha+\frac{\bar\beta}\beta\ln\bar\beta\Big],
\notag\\
\varphi_3(\alpha,\beta)&=-\ln(1-\alpha-\beta)\theta(1-\tau)\,.
\end{align}
They are defined up to arbitrary solutions of the corresponding homogeneous equations: a constant for $\phi_{1,3}(\alpha)$
and a function of the conformal ratio for $\varphi_{1,3}(\alpha,\beta)$.
These missing pieces and also the complete kernel $h^{(2)}_{2}(\tau)$ can be fixed from the known spectrum of the two-loop
anomalous dimensions.

The well-known decomposition
\begin{align}
\gamma^{(2)}_N & = m^+_N + (-1)^{N-1} m_N^-\,,
\end{align}
where $m^\pm_N$ can be extended to analytic functions $m^\pm(N)$ with poles at negative real axis,
corresponds for light-ray operators to the decomposition in integration regions where the quark
and the antiquark retain their ordering on the light cone or are interchanged
\begin{align}
h(\tau)=\theta(1-\tau)h^+(\tau)+\theta(\tau-1)h^-(\tau^{-1})\,,
\end{align}
corresponding to $\alpha+\beta < 1$ and $\alpha+\beta > 1$, respectively.
The equations
\begin{align}\label{mN}
  m ^\pm(N) = \int_0^1d\alpha\int_0^{\bar \alpha }d\beta\,h^\pm(\tau) \,(1-\alpha-\beta)^{N-1}
\end{align}
can be inverted as
\begin{align}
h^\pm (\tau)=\frac{1}{2\pi i}\int_{-i\infty}^{+i\infty} dN (2N+1)\, m^\pm (N)\, P_{N}\left(\frac{1+\tau}{1-\tau}\right),
\end{align}
where $P_N$ is the Legendre function.
The integration goes along the imaginary axis such that all poles of $m^{\pm}(N)$ 
lie to the left of the integration contour.
In practice it turns out to be more efficient to start from a certain ``educated guess''  for the kernels,
calculate the moments and find the coefficients.

The final result reads
\begin{align}
h^{(2)}_1(\alpha,\beta)&=-\delta_+(\tau)\Big(\phi_1(\alpha)+\phi_1(\beta)\Big)+\varphi_1(\alpha,\beta)
+ \theta(\bar\tau)\left[2\Li_2(\tau)+\ln^2\bar\tau+\ln\tau-\frac{1+\bar\tau}{\tau}\ln\bar\tau\right]
\notag\\
&\quad+
\theta(-\bar\tau)\left[\ln^2(-\bar\tau/\tau)-\frac2\tau\ln(-\bar\tau/\tau)\right]
+\left(-6\zeta(3)+\frac13\pi^2+\frac{21}8\right)\delta(\alpha)\delta(\beta)\,,
\notag\\
h^{(2)}_2(\alpha,\beta)&=\frac13\left({\pi^2}-4\right)\delta_+(\tau) - 2\theta(\bar\tau)\left[\Li_2(\tau)-\Li_2(1)+\frac12\ln^2\bar\tau
-\frac1\tau\ln\bar\tau+\frac53\right]
\notag\\
&\quad -\theta(-\bar\tau)\left[\ln^2(-\bar\tau/\tau)-\frac2\tau\ln(-\bar\tau/\tau)\right]
+ \left(6\zeta(3)-\frac23\pi^2+\frac{13}6\right)\delta(\alpha)\delta(\beta)\,,
\notag\\
h^{(2)}_3(\alpha,\beta)&=-\delta_+(\tau)\left[\ln\bar\alpha+\ln\bar\beta+\frac53\right]-\theta(\bar\tau)
\left[\ln(1-\alpha-\beta)+\frac{11}3\right]+\frac{13}{12}\delta(\alpha)\delta(\beta)\,,
\label{final}
\end{align}
where $\bar\tau = 1-\tau$, and the functions $\phi_1(\alpha)$ and $\varphi_1(\alpha,\beta)$ are given in Eq.~(\ref{phi1}).

%
\vspace*{0.5cm}
%

{\large\bf 6.}~~
Our result for the two-loop evolution of flavor-nonsinglet light-ray operators in Eqs.~(\ref{CACFH}), (\ref{final})
is equivalent to the corresponding evolution equation  for GPDs obtained
in Ref.~\cite{Belitsky:1999hf} but is more compact and has manifest $SL(2)$ symmetry properties.
The latter feature presents the crucial advantage of the
light-ray operator (alias position space) representation which makes this technique attractive for
higher-order calculations. Exact conformal symmetry of QCD at the critical point proves to be very helpful 
as it provides one with algebraic group-theory methods to calculate
the commutators of integral operators that appear in Eqs.~(\ref{nest}).
Evolution equations for GPDs can be obtained from our expressions by a Fourier
transformation which is rather straightforward, cf.~\cite{Braun:2009mi}.

Apart from the evolution kernels, another new result is the calculation
of the generator of special conformal transformations $S_{+}$ to the
one-loop accuracy, see Eq.~(\ref{exactS}). The QCD expression differs from the corresponding
result in the scalar theory~\cite{Braun:2013tva} but remains simple.
As we have demonstrated, this result can be obtained from the gauge-invariant correlation function
of two light-ray operators, thus bypassing the complications due to non-gauge-invariant contributions
in the usual approach dealing with Green functions involving fundamental fields.
We expect that the same technique can be used for the flavor-singlet light-ray operators and for
the calculation of $S_{+}$ to the two-loop accuracy, which is the first step towards
the NNLO evolution equations. This task, obviously, goes beyond the scope of this Letter.


\vskip5mm

{\large \bf 7.}~~{\large\bf Acknowledgments}\\
We thank D.~Mueller for useful discussions.
This work was supported by the DFG, grant BR2021/5-2.


%

\end{document}